\documentclass[12pt,aps,preprint,floats,floatfix,amsmath,nofootinbib,superscriptaddress]{revtex4-1}
\usepackage{setspace}
\usepackage{graphicx}
\usepackage{bm}
\usepackage{epsfig}
\usepackage{latexsym}
\usepackage{color}
\usepackage{colordvi}
\usepackage{verbatim}
\usepackage{amssymb}

\newcommand{\beq}{\begin{equation}}
\newcommand{\eeq}{\end{equation}}
\newcommand{\bea}{\begin{eqnarray}}
\newcommand{\eea}{\end{eqnarray}}

\begin{document}

\title{Superembedding Methods for Current Superfields}

\author{Walter D. Goldberger, Zuhair U. Khandker, Daliang Li, and Witold Skiba}
\affiliation{Department of Physics, Yale University, New Haven, CT 06520}

\begin{abstract}
We extend the superembedding formalism for 4D $\mathcal{N}=1$ superconformal field theory (SCFT)  to the case of fields in arbitrary representations of the superconformal group $SU(2,2|1)$.  As applications we obtain manifestly superconformally covariant expressions for two- and three-point functions involving conserved currents, e.g. the supercurrent multiplet or global symmetry current superfields.   The embedding space results are presented in a compact form by employing an index-free formalism.   Our expressions are consistent with the literature, but the manifestly covariant forms of correlators presented here are new.  
\end{abstract}

\maketitle

\section{Introduction}
\label{Introduction}

Four-dimensional conformal symmetry imposes stringent constraints on the form of quantum field theory correlators, as well as restrictions on scaling dimensions of certain operators~\cite{CFT}. However, the full implications of conformal invariance in four dimensions are not yet completely known.  Recently, there has been some progress in using general principles such as unitarity, the operator product expansion (OPE) and its conformal block decomposition, and crossing symmetry to derive constraints on four-dimensional conformal field theories (CFTs).   See for instance Refs.~\cite{Rattazzi:2008pe} for recent results in this direction.

A useful tool for exploring the consequences of conformal invariance in 4D is the embedding space formalism~\cite{Dirac:1936fq,Mack:1969rr}, in which four-dimensional Minkowski space is identified with the projective lightcone in a flat six-dimensional space with signature $(4,2)$ metric, which we will refer to as the `embedding space'. The conformal group $SO(4,2)$ acts linearly on the embedding space, so working in this framework makes conformal symmetry manifest at the level of the correlation functions.  The embedding space language was first applied to field theory in Refs.~\cite{Dirac:1936fq} to derive manifestly covariant free wave equations, and employed in the context of general interacting CFTs in~\cite{Mack:1969rr}. 
 From the point of view of constraining CFTs, applications of embedding space methods include a conformally covariant formulation of the OPE~\cite{Ferrara:1973yt}, and the derivation of closed-form expressions for the conformal partial wave decomposition of four-point functions in four and six dimensions~\cite{Dolan:2003hv}.    More recent results can be found, e.g., in~\cite{Costa:2011mg,SimmonsDuffin:2012uy}.

In this paper, we consider a supersymmetric version of the projective lightcone formalism which is appropriate to describe 4D ${\cal N}=1$ superconformally invariant field theory (SCFT), following the recent work of Ref.~\cite{GSS}.   One motivation for focusing on supersymmetric conformal theories is that they provide a large sample of interacting CFTs with often tractable dynamics which could be used to explicitly test the recent ideas discussed in Refs.~\cite{Rattazzi:2008pe}.    Indeed, most non-trivial 4D CFTs with a known microscopic realization (apart from perturbative Banks-Zaks type fixed point theories)  are in fact SCFTs.

Ref.~\cite{GSS} showed how to realize the $SU(2,2|1)$ ${\cal N}=1$ superconformal symmetry on an embedding superspace whose coordinates transform linearly under $SU(2,2|1)$.   These coordinates are spanned by a set of supermatrices that decompose into seven bosonic and four Grassmann components,  and transform in eleven-dimensional irreducible representations of $SU(2,2|1)$.  
Four-dimensional Minkowski superspace is realized in terms of a set of covariant quadratic constraints on these coordinates.    Related work on supersymmetric generalizations of embedding space methods include~\cite{Ferrara:1974qk}  which makes $SO(4,2)$ rather than  $SU(2,2|1)$ invariance manifest, and Refs.~\cite{Siegel:1994cc,Kuzenko:2006mv} that employ supertwistor techniques.    The ${\cal N}>1$ extension of the construction discussed in~\cite{GSS} was obtained recently in~\cite{Maio:2012tx}, see also Ref.~\cite{Kuzenko:2012tb}.  

Here, we further develop the realization of superconformal fields in the language of Ref.~\cite{GSS}.   In particular, we establish a correspondence between superfields of arbitrary spin on Minkowski superspace and superfields in the embedding space and use it to work out the implications for two-point and three-point correlators.     To illustrate our methods, we focus on the physically relevant cases of global symmetry current multiplets, described in four dimensions by real scalar superfields, and the supercurrent multiplet.   Our results are presented in a compact index-free notation analogous to the one developed in~\cite{Costa:2011mg,SimmonsDuffin:2012uy} for non-supersymmetric CFTs.   When written in four-dimensional language, our results agree with the existing literature~\cite{Molotkov:1975gf,Osborn:1998qu,Park:1997bq}.

The paper is organized as follows. In the next section, we review the superembedding formalism of Ref.~\cite{GSS} and establish our notation. In Sec.~\ref{Dictionary}, we establish a correspondence between superfields in arbitrary Lorentz representations and their superembedding space counterparts. In Sec.~\ref{Correlators}, we apply our formalism to examples of 2-point and 3-point correlators.  Emphasis is placed on correlators involving the various symmetry currents of the SCFT.   In particular, we recover the known superconformal relations between possible anomalies of global currents. Construction of manifestly covariant correlators reduces to the problem of enumerating $SU(2,2|1)$ invariants built from products of embedding space supercoordinates with a fixed number of super-twistors and their complex conjugates.    We conclude in Sec.~\ref{Conclusion}.

\section{Superembedding Formalism}
\label{Superembedding Formalism}

In order to establish our notation we briefly review the  `superembedding formalism' developed in Ref.~\cite{GSS}. The four-dimensional $\mathcal{N}=1$ Minkowski superspace, $\mathcal{M}=\mathbb{R}^{4|4}$, is embedded in a higher-dimensional superspace $\mathcal{E}$ on which the superconformal group $SU(2,2|1)$ acts linearly. The reduction from the embedding space  $\mathcal{E}$ to the four-dimensional superspace $\mathcal{M}$ is accomplished by a set of covariant constraints, which we review in this section.

\subsection{The Superconformal Group}

Our notation for the  $\mathcal{N}=1$ superconformal group $SU(2,2|1)$ follows that of Ref.~\cite{GSS}.    The supergroup $SU(2,2|1)$  consists $5\times 5$ supermatrices\footnote{Capital indices $A,B,\ldots$ run over $A=\alpha=1,\ldots,4$ or $A=5$} of the form 
\begin{equation}
U_A^{\phantom{A}B} = \left(
\begin{array}{cc}
U_\alpha^{\phantom{\alpha}\beta} & \phi_\alpha \\
\psi^\beta & z
\end{array} \right),
\label{UAB}
\end{equation}
with bosonic ($c$-number) entries  $U_\alpha^{\phantom{\alpha}\beta}$ and $z$, and (anticommuting) fermionic entries  $\phi_\alpha$ and $\psi^\beta$.    These matrices act on a fundamental (defining) five-dimensional representation $V_A$, where we assign $V_\alpha$ to be fermionic and $V_5$ is bosonic, as $V_A\rightarrow U_A^{\phantom{A}B}  V_B$.   On the conjugate representation $V_{\dot A} \equiv V_A^\dagger$, $SU(2,2|1)$ acts as $V_{\dot A}\rightarrow V_{\dot B} U_{\phantom{\dot B}\dot A}^{\dot B}$, with $U_{\phantom{\dot B}\dot A}^{\dot B}=\left(U_A^{\phantom{A}B}\right)^\dagger$.  A supermatrix $U_A^{\phantom{A}B}$  belongs in $SU(2,2|1)$ if it satisfies the ``unitarity" constraint
\begin{equation}
A^{\dot{A}A} = U_{\phantom{A} \dot{B}}^{\dot{A}} A^{\dot{B}B} U_B^{\phantom{A}A}, 
\label{MetricInvt}
\end{equation}
where the $SU(2,2|1)$ invariant metric is given by
\begin{equation}
A^{\dot{A}B} = \left( 
\begin{array}{cc}
A^{\dot{\alpha}\beta} & 0 \\
0 & 1
\end{array} \right)
= \left(
\begin{array}{ccc}
0 & \delta^{\dot{a}}_{\phantom{a} \dot{b}} & 0 \\
\delta_a^{\phantom{a}  b} & 0 & 0 \\
0 & 0 & 1
\end{array}\right),
\label{Metric}
\end{equation}
as well as the ``unimodular" constraint
\begin{equation}
\left[\mathrm{sdet}\, U\right]^{-1} = \frac{\mathrm{det}\left(U_\alpha^{\phantom{\alpha}\beta} - z^{-1}\phi_\alpha\psi^\beta\right)}{z} = 1.
\label{Sdet}
\end{equation}
 
It is most often convenient to work with infinitesimal generators rather than with finite group elements.   Near the identity,
\begin{equation}
U_A^{\phantom{A}B} = \delta_A^{\phantom{A}B} + i\, T_A^{\phantom{A}B},
\end{equation}
where the generators take the form
\begin{equation}
T_A^{\phantom{A}B} = \left(
\begin{array}{cc}
T_\alpha^{\phantom{\alpha}\beta} + \frac{1}{4}\delta_\alpha^{\phantom{\alpha}\beta} \phi & \phi_\alpha \\
\bar{\phi}^\beta & \phi
\end{array} \right).
\label{TAB}
\end{equation}
The traceless tensor  $T_\alpha^{\phantom{\alpha}\beta}$ is a generator of $SU(2,2)\subset SU(2,2|1)$ (the invariant $SU(2,2)$ metric is the matrix $A^{\dot{\alpha} \beta}$ defined above)  and $\bar{\phi}^\beta \equiv  \phi_{\dot{\alpha}} A^{\dot{\alpha} \beta}$.     The action of the generators on fundamental  and anti-fundamental representations is then
\begin{equation}
\delta V_A = i T_A^{\phantom{A}B} V_B,
\label{FundamentalTransf}
\end{equation}
and
\begin{equation}
\delta \bar{V}^A = -i \bar{V}^B T_B^{\phantom{A}A},
\label{AntiFundamentalTransf}
\end{equation}
where we have introduced the notation $\bar{V}^A \equiv V_{\dot{A}} A^{\dot{A} A}$.

Tensor products of fundamental and antifundamental $V_A$ and $\bar{V}^A$ representations generally have mixed symmetry properties.   To keep track of minus signs that arise when permuting such objects, we employ the notation
\begin{eqnarray}
\label{eq:sigma}
\sigma(AB) &\equiv& \left\{\begin{array}{cl}
-1 & \text{if  both $A$ and $B$ are fermionic indices $\alpha,\beta$},\\
+1 & \text{otherwise,}
\end{array}\right. \\
\sigma(A) &\equiv& \sigma(AA).
\end{eqnarray}
For instance,
\begin{eqnarray}
V_A W_B &=& \sigma(AB) W_B V_A, \label{eq:example1} \\
V_A T_B^{\phantom{B}C} &=& \sigma(AB)\sigma(AC) T_B^{\phantom{B}C} V_A.  \label{eq:example2} 
\end{eqnarray}
Note that the indices inside $\sigma()$ do not obey the standard repeated index sum convention, i.e. no sum is implied on the right-hand sides of Eqs.~(\ref{eq:example1}) and (\ref{eq:example2}). However, when an index is repeated, not including the arguments of $\sigma()$'s in the count, then a single sum is implicit over such an index.  Eqs.~(\ref{FundamentalTransf}) and (\ref{AntiFundamentalTransf}) imply that 
\begin{equation}
 W_A \bar{V}^A \sigma(A)=\bar{V}^A W_A 
\label{Invt}
\end{equation}
is an $SU(2,2|1)$ singlet, and as we already mentioned a sum over the index $A$ is implicit on both sides.  This is a specific case of a general rule for covariantly contracting $SU(2,2|1)$ indices when $\alpha$ is fermionic and $5$ is bosonic.  From left to right, an upper index contracts with a lower index without any $\sigma$-factor, and from left to right, a lower index contracts with an upper index with a $\sigma$-factor.   Extensive discussions of $SU(2,2|\mathcal{N})$ representations can be found in Refs.~\cite{Flato:1983te,Dobrev:1985qv,Minwalla:1997ka}, and a discussion of tensor product representations and super Young tableaux can be found in Refs.~\cite{Baha Balantekin:1980qy,Bars:1984rb}, although for our purposes, the properties summarized above will suffice.

\subsection{Superembedding Space}
\label{SuperembeddingSpace}

We introduce an embedding superspace, which we will refer to as the superembedding space $\mathcal{E}$, that contains four-dimensional superspace ${\mathbb R}^{4|4}$ (denoted by  ${\cal M}$) and transforms linearly under superconformal transformations.   It is  the analog of the embedding space for 4D CFTs ${\mathbb R}^6$  with coordinates $X^m$ (with $m$ taking the values $+$, $-$ or, $\mu=0,1,2,3$) that transforms linearly under $SO(4,2)$ and contains the conformal  compactification of four-dimensional Minkowski space, realized as the projective lightcone $0=X^2 = \eta_{mn} X^m X^n =  \eta_{\mu\nu} X^\mu X^\nu + X^+ X^-$~\cite{Dirac:1936fq,Mack:1969rr}.

To do so, we need a `coordinate supermultiplet' $X_{AB}$ defined to possess identical $SU(2,2|1)$ transfomation properties and index exchange symmetry as the tensor product $V_A V_B$,  i.e.
\begin{eqnarray}
X_{AB} &=& \sigma(AB) X_{BA},
\end{eqnarray}
with infinitesimal $SU(2,2|1)$ transformation
\begin{equation}
\delta X_{AB} = i T_A^{\phantom{A}B'} X_{B'B} + i\sigma(AB) T_B^{\phantom{B}A'} X_{A'A}.
\label{deltaX}
\end{equation}
The multiplet $X_{AB}$ contains the bosonic components $X_{\alpha\beta}=-X_{\beta\alpha}$,  $X_{55}\equiv \varphi$ and fermionic coordinates $X_{5\alpha}=X_{\alpha5}\equiv \theta_\alpha$.   The anti-symmetric $SU(2,2)$ tensor $X_{\alpha\beta}$ can be equivalently written in $SO(4,2)$ notation as the six-dimensional vector $X^m$.   The explicit correspondence is
\begin{equation}
X^m = \frac{1}{2}X_{\alpha\beta} \Gamma^{m\,\alpha\beta}, \hspace{10mm} X_{\alpha\beta} = \frac{1}{2}X_m \tilde{\Gamma}^m_{\alpha\beta},
\label{Map}
\end{equation}
where the matrices $\Gamma^{m\,\alpha\beta}$ and $\tilde{\Gamma}^m_{\alpha\beta}$ and their properties are given in Appendix A of Ref.~\cite{GSS}.

Because there is no covariant reality condition in $SU(2,2|1)$, $X_{AB}$ is in a complex representation.    We introduce an additional coordinate ${\bar X}^{AB}$ with the same properties as the tensor product ${\bar V}^A {\bar V}^B$, i.e. ${\bar X}^{AB} =\sigma(AB) \bar{X}^{BA},$ and 
\begin{equation}
\delta \bar{X}^{AB} = -i \bar{X}^{AA'}T_{A'}^{\phantom{A}B} - i\sigma(AB) \bar{X}^{BB'}T_{B'}^{\phantom{B}A}.
\label{deltaXbar}
\end{equation}
The superembedding space $\mathcal{E}$ consists of the space ${\mathbb C}^{7|4}$ spanned by the pair $\left(X_{AB},\bar{X}^{AB}\right)$.

The real four-dimensional Minkowski superspace $\mathcal{M}$ is recovered as the subset of $\mathcal{E}$ obtained by projective identification $\left(X,\bar{X}\right) \sim \left(\lambda X,\bar{\lambda} \bar{X}\right)$, and by imposing the relation
\begin{equation}
\bar{X}^{AB} = A^{\dot{A}A}A^{\dot{B}B} X_{\dot{A}\dot{B}}, \hspace{5mm} \mathrm{where} \hspace{5mm} X_{\dot{A}\dot{B}} \equiv \left( X_{BA} \right)^\dagger,
\label{XbarDefn}
\end{equation}
between $X_{AB}$ and ${\bar X}^{AB}$, together with the following constraints~\cite{GSS,MS}:
\begin{eqnarray}
  \left[X_{AB} X_{CD}\right]_{\bf{16}} &=& 0, \ \ \  \left[\bar{X}^{AB} \bar{X}^{CD}\right]_{\bf{\overline{16}}}=0,   \label{constraints16} \\
  \left[ X_{AB} \bar{X}^{BC} \right]_{\bf{24}} &=& 0, \label{constraints24}
\end{eqnarray}
where the boldface subscripts denote the dimensions of the irreducible $SU(2,2|1)$ representation that we project onto. For example, the (adjoint) ${\bf 24}$ representation  consists of supermatrices $M_A{}^B$ with zero supertrace, 
\begin{equation}
\mbox{str} M \equiv - \sigma(A) M_A{}^A=0,
\label{eq:str}
\end{equation}
while $\left[X_{AB} X_{CD}\right]_{\bf{16}} =0$ is equivalent to the cyclic constraint:
\begin{equation}
X_{\underline{A}\underline{B}} X_{\underline{C}D} \equiv X_{AB}X_{CD} + \sigma(AC)\sigma(AB)X_{BC}X_{AD} + \sigma(AC)\sigma(BC)X_{CA}X_{BD} = 0,
\label{16}
\end{equation}
where appropriate $\sigma$-factors are inserted to ensure $SU(2,2|1)$-covariance.   Solutions of the constraint Eqs.~(\ref{constraints16}) and (\ref{constraints24}) also automatically satisfy $[X_{AB} \bar{X}^{AB}]_{\bf 1}=0$. Therefore, $X_{AB} \bar{X}^{BC}=0$ for any values of $A$ and $C$.

To see that four-dimensional superspace $\mathcal{M}$ corresponds to the subspace of $\mathcal{E}$ defined by these equations, we note that solutions can be generated, at least locally, by applying all possible $SU(2,2|1)$ to any single point obeying the constraints~\cite{GSS}.     For example, one may start from `the origin'
\begin{equation}
\hat{X}_{AB} = \left(
\begin{array}{ccc}
\frac{i}{2}\epsilon_{ab} X^+ & 0 & 0 \\
0 & 0 & 0 \\
0 & 0 & 0
\end{array}\right),
\label{Origin}
\end{equation}
which is (projectively) invariant under $SO(3,1)\subset SU(2,2|1)$, special conformal transformations, special superconformal transformations, dilatations, and by a global $U(1)_R\subset SU(2,2|1)$.    It follows that the space of solutions to the constraints can be identified with the coset  $SU(2,2|1)/H$, with $H$ the isotropy group of the origin.   This $(4+4)$-dimensional space is Minkowski superspace ${\cal M}$.   A convenient parameterization near the origin is $\left(X^+,\bar{X}^+,x^\mu,\theta_a,\bar{\theta}^{\dot{a}}\right)$, where $a,{\dot a}=1,2$ are two-component $SL(2,{\mathbb C})$ indices (our conventions for two-component spinors are those of Wess and Bagger~\cite{Wess:1992cp}), and:
\begin{equation}
X_{AB} = X^+ \left(
\begin{array}{ccc}
\frac{i}{2}\epsilon_{ab} & \frac{1}{2}\left(y\sigma\epsilon\right)_a^{\phantom{a}\dot{b}} & \theta_a \\
-\frac{1}{2}\left(y \bar{\sigma}\epsilon\right)^{\dot{a}}_{\phantom{a}b} & -\frac{i}{2}y^2\epsilon^{\dot{a}\dot{b}} & i\left(y\bar{\sigma}\theta\right)^{\dot{a}}  \\
\theta_b & i\left(y\bar{\sigma}\theta\right)^{\dot{b}}  & 2i\theta^2
\end{array}\right),
\label{XParametrize}
\end{equation} 
\begin{equation}
\bar{X}^{AB} = \bar{X}^+ \left(
\begin{array}{ccc}
-\frac{i}{2}\bar{y}^2\epsilon^{ab} & -\frac{1}{2}\left(\epsilon\sigma \bar{y}\right)^{a}_{\phantom{a}\dot{b}} & \; -i\left(\bar{\theta}\bar{\sigma} \bar{y}\right)^{a}  \\
\frac{1}{2}\left(\epsilon\bar{\sigma} \bar{y}\right)_{\dot{a}}^{\phantom{b}b} & \frac{i}{2}\epsilon_{\dot{a}\dot{b}} & \bar{\theta}_{\dot{a}} \\
-i\left(\bar{\theta}\bar{\sigma} \bar{y}\right)^{b} & \bar{\theta}_{\dot{b}} & -2i\bar{\theta}^2
\end{array} \right).
\label{XbarParametrize}
\end{equation}
The four-dimensional coordinates satisfy  $y^\mu-{\bar y}^\mu = 2 i \theta \sigma^\mu {\bar\theta}$ on account of the $[X \bar{X}]_{\bf{24}}=0$ constraint, and $x^\mu=\frac{1}{2} (y^\mu+ \bar{y}^\mu)$.    It is straightforward to verify~\cite{GSS} that $(x^\mu,\theta_a, {\bar\theta}^{\dot a}$) transform in the standard way under superconformal transformations (for instance in the form given in~\cite{Buchbinder:1998qv}).   Note that the the upper $4\times 4$ block in Eq.~(\ref{XParametrize}) is $\frac{1}{2} X_m {{\tilde\Gamma}^m}_{\alpha\beta}$ with $X^m = \left( X^+, X^\mu = X^+ y^\mu, X^- = -X^+ y^2 \right)$ on the (complexified) $SO(4,2)$ lightcone.   Similar results hold for the conjugate coordinates.

When $X_{AB}$ is restricted to $\mathcal{M}$ as in Eq.~(\ref{XParametrize}) its components are linearly dependent. The second and third columns of the matrix $X_{AB}$ can be given in terms of the first column
\begin{equation}
  X_A^{\phantom{A}\dot{b}}= i y \overline{\sigma}^{\dot{b} b} X_{Ab} \ \ {\rm and} \ \ X_{A5}=2 i \theta^b X_{Ab} \, \sigma(A). \label{eq:Xcomponents}
\end{equation}
Similar relations for $\bar{X}^{AB}$ relate the first and third rows to the second one:
\begin{equation}
  \bar{X}^{bA}= - i  \bar{X}_{\dot{b}}^{\phantom{b} A}  \overline{\sigma}^{\dot{b} b} \ \ {\rm and} \ \ \bar{X}^{5A}=- 2 i \bar{X}_{\dot{b}}^{\phantom{b} A}  \overline{\theta}^{\dot{b}} \sigma(A).
\end{equation}

The parametrization of ${\cal M}$ given above describes points near the origin $x^\mu=0$.    Points near infinity, with $z^\mu = x^\mu/x^2$ close to $z^\mu=0$, can be described by applying all possible $SU(2,2|1)$ transformations to 
\begin{equation}
\check{X}_{AB} = \left(
\begin{array}{ccc}
0 & 0 & 0 \\
0 & \frac{i}{2}\epsilon_{\dot{a}\dot{b}} X^- & 0 \\
0 & 0 & 0
\end{array}\right),
\label{Infinity}
\end{equation}
which is left invariant by $SL(2,{\mathbb C})$  Lorentz transformations, translations and Poincare supersymmetry transformations.   Given two points in $\mathcal{M}$ there exists an $SU(2,2|1)$ transformation that simultaneously sends one of the points to the `origin', Eq.~(\ref{Origin}), and the other to `infinity' given in Eq.~(\ref{Infinity}).

\section{Superfields}
\label{Dictionary}

We now develop the correspondence between superfields on $\mathcal{M}$ and their counterparts in $\mathcal{E}$.   A generic  primary superfield $\Phi_{\mathcal{M}}$ on $\mathcal{M}$ is specified by its $SL(2,{\mathbb C})$ Lorentz quantum numbers, by its scaling dimension $\Delta$ and by the $U(1)_R$ charge of its lowest component field~\cite{Osborn:1998qu}.    It is given by an $SL(2,{\mathbb C})$ multi-spinor,             
\begin{equation}
\Phi_{\mathcal{M}\,a_1\cdots a_{2j}}{}^{\dot{b}_1\cdots \dot{b}_{2{\bar j}}} \left(x,\theta,\bar{\theta}\right),
\end{equation}
where irreducibility requires complete symmetry under the interchange of pairs of dotted or undotted indices.     It is useful to label this superfield by its quantum numbers  $\left(j,\bar{j},q,\bar{q}\right)$, where 
\begin{eqnarray}
q &\equiv& \frac{1}{2}\left(\Delta + \frac{3}{2} R\right), \label{q} \\
\bar{q} &\equiv& \frac{1}{2}\left(\Delta - \frac{3}{2} R\right). \label{barq}
\end{eqnarray}
The correspondence for  $j=\bar{j}=0$ was developed in Ref.~\cite{GSS}:     A $(0,0,q,{\bar q})$ primary operator $\Phi_{\cal M}(x,\theta,{\bar\theta})$ in Minkowski spacetime maps into an embedding space $SU(2,2|1)$ scalar superfield $\Phi_{\cal E}(X,{\bar X}) = \left(X^+\right)^{-q} \left({\bar X}^+\right)^{-{\bar q}} \Phi_M(x,\theta,{\bar\theta})$ homogeneous in its arguments.    We now extend this result to other representations.

First, we recall the mapping between primary operators of $SO(4,2)$ in spinor representations and their projective lightcone countertparts.  Recent, detailed discussions can be found in refs.~\cite{Weinberg:2010fx,SimmonsDuffin:2012uy}.     Starting with  primary $\psi_a(x^\mu)$ transforming in the $(1/2,0)$ representation of  $SL(2,{\mathbb C})$,  one constructs a projective lightcone field 
\begin{equation}
\psi_\alpha(X) = \left(X^+\right)^{-\Delta-1/2} X_{\alpha b} \psi^b(x^\mu),
\label{eq:4to6spinor}
\end{equation}
which is a homogeneous function, $\psi_\alpha(\lambda X) = \lambda^{-\Delta+1/2} \psi_\alpha(X),$  transforming in the fundamental representation of $SU(2,2)\sim SO(4,2)$.   Points on the projective lightcone automatically obey the stronger constraint $X_{\alpha\lambda} X^{\lambda\beta}=0$ and thus $\psi_\alpha(X)$ defined by Eq.~(\ref{eq:4to6spinor}) satisfies the relation 
\begin{equation}
X^{\alpha\beta} \psi_\beta(X)=0.
\label{eq:6spincon}
\end{equation}
 Conversely, given a spinor $\psi_\alpha(X)$ on the projective lightcone satisfying the constraint in Eq.~(\ref{eq:6spincon}), it is possible to project onto a spinor primary field $\psi_a(x^\mu)$ in Minkowski space,\footnote{Note that there is some redundancy in this construction, as the lightcone constraint on $X_{\alpha\beta}$ means that the solution to Eq.~(\ref{eq:6spincon}) is not unique.   Given any solution to Eq.~(\ref{eq:6spincon}), one may generate others by adding terms of the form $X_{\alpha\beta} \chi^\beta(X)$.    While this freedom changes the explicit expression for $\psi_a(x^\mu)$ in Eq.~(\ref{eq:6to4spinor}), it does not change results for correlation functions obtained via the embedding space formalism, see~\cite{SimmonsDuffin:2012uy} for details.}
\begin{equation}
\psi_a(x^\mu) = \left(X^+\right)^{\Delta-1/2} \psi_{\alpha=a}(X).
\label{eq:6to4spinor}
\end{equation}
This correspondence generalizes to fields in other representations in the obvious way.

In order to establish the analogous correspondence for superconformal fields, we need to supersymmetrize the constraint in Eq.~(\ref{eq:6spincon}).     The generalization of $\psi_\alpha(X)$ is an embedding space superfield $\Phi_{\mathcal{E}\,A}(X,{\bar X})$ in the fundamental representation of $SU(2,2|1)$ satisfying the scaling property
\begin{equation}
\Phi_{\mathcal{E}\,A}\left(\lambda X,\bar{\lambda} \bar{X}\right) = \lambda^{-\left(q-\frac{1}{2}\right)} \bar{\lambda}^{-\bar{q}} \Phi_{\mathcal{E}\,A}\left(X,\bar{X}\right).
\label{UndottedHomog}
\end{equation}
To generalize Eq.~(\ref{eq:6spincon}) we must find a supermultiplet of linear constraints constructed from the product of  $\Phi_{\mathcal{E}\,A}$ with either $X_{AB}$ or ${\bar X}^{AB}$.   The two possibilities are either ${\bar X}^{AB}  \Phi_{\mathcal{E}\,B}=0$, which reduces to Eq.~(\ref{eq:6spincon}) at the point ${\bar X}^{5\alpha}={\bar X}^{55}=0$ or  $[X_{AB} \Phi_{{\cal E}\,C}]_{\bf 15}=0$, which can be written explicitly as
\begin{equation}
X_{\underline{A}\underline{B}} \Phi_{\mathcal{E}\,\underline{C}} \left(X,\bar{X}\right) \equiv X_{AB}\Phi_{\mathcal{E}\,C} + \sigma(AC)\sigma(AB)X_{BC}\Phi_{\mathcal{E}\,A} + \sigma(AC)\sigma(BC)X_{CA}\Phi_{\mathcal{E}\,B} = 0,
\label{eq:15phicon}
\end{equation}
or component-wise
\begin{eqnarray}
 \label{eq:con4barof15} 
X^{\alpha \beta} \Phi_{\mathcal{E}\,\beta} &=& 0, \\
\label{eq:con6of15} 
 X_{\alpha \beta} \Phi_{\mathcal{E}\,5} + \theta_\alpha \Phi_{\mathcal{E}\,\beta} - \theta_\beta \Phi_{\mathcal{E}\,\alpha} &=& 0,  \\
      2 \theta_\alpha \Phi_{\mathcal{E}\,5} + \varphi  \Phi_{\mathcal{E}\,\alpha} &=& 0, \label{eq:con4of15} \\
   \varphi  \Phi_{\mathcal{E}\,5} &=& 0. \label{eq:con1of15} 
\end{eqnarray}
In particular, the $A=\alpha$, $B=\beta$, $C=\gamma$ component of Eq.~(\ref{eq:15phicon}), given in Eq.~(\ref{eq:con4barof15}), implies that $\Phi_{\mathcal{E}\,\alpha}$ obeys Eq.~(\ref{eq:6spincon}) at all points of ${\cal E}$.  Furthermore it can be checked that imposing $[X_{AB} \Phi_{{\cal E}\,C}]_{\bf 15}=0$ on ${\cal E}$ also implies that $\Phi_{\mathcal{E}\,A}$ obeys the anti-fundamental constraint ${\bar X}^{AB}  \Phi_{\mathcal{E}\,B}=0$ restricted to ${\cal M}$.

From these observations, we conclude that $[X_{AB} \Phi_{{\cal E}\,C}]_{\bf 15}=0$ yields the constraint necessary to recover the correct four-dimensional superfield $\Phi_{{\cal M}, a}(x^\mu,\theta,{\bar\theta})$.   Indeed, for fixed $\Phi_{\mathcal{E}\,\alpha=a}$ and $X_{AB}$ restricted to $\mathcal{M}$, using the parametrization in Eq.~(\ref{XParametrize}), Eq.~(\ref{eq:con4of15}) implies 
\begin{equation}
 \Phi_{\mathcal{E}\,5} =  2 i \theta^a \Phi_{\mathcal{E}\,a}, \ \ \ 
 \Phi_{\mathcal{E}}^{\dot{a}} = i y^\mu\, \overline{\sigma}_\mu^{\dot{a} a} \, \Phi_{\mathcal{E}\,a}.
 \label{UndottedConstraintComponents} 
\end{equation}
The remaining constraints, Eqs.~(\ref{eq:con4barof15}),  (\ref{eq:con6of15}), and (\ref{eq:con1of15}), are then automatically satisfied given Eq.~(\ref{UndottedConstraintComponents}).

Since only the $A=a$ components of $\Phi_{\mathcal{E}\,A}$ are left independent by the constraint in Eq.~(\ref{eq:15phicon}), we define $\Phi_{\mathcal{M}\,a}(x^\mu,\theta,{\bar\theta})$ as 
\begin{equation}
\Phi_{\mathcal{M}\,a} \left(x,\theta,\bar{\theta}\right) = \left(X^+\right)^{q-\frac{1}{2}} \left(\bar{X}^+\right)^{\bar{q}} \Phi_{\mathcal{E}\,A=a} \left(X,\bar{X}\right),
\label{UndottedProjection} 
\end{equation}
where the coordinates on the right-hand side are restricted to $\mathcal{M}$. It is also possible to uplift the superfield $\Phi_{\mathcal{M}\,a}$ from ${\cal M}$ to ${\cal E}$ via
\begin{equation}
  \Phi_{\mathcal{E}\, A}(X,\bar{X})=\left(X^{+}\right)^{-(q+\frac{1}{2})}\left(\bar{X}^{+}\right)^{-\bar{q}}X_{A}^{\phantom{A}c}\Phi_{\mathcal{M} \, c}(x,\theta,\bar{\theta}),
  \label{UndottedInverseProjection} 
\end{equation}
which follows by using Eqs.~(\ref{UndottedConstraintComponents}), (\ref{UndottedProjection}), and (\ref{eq:Xcomponents}).   Of course, it is understood that the field $\Phi_{\mathcal{E}\, A}(X,\bar{X})$ is not defined for all values of $X$ and $\bar{X}$ in the superembedding space $\mathcal{E}$ since $\Phi_{\mathcal{M} \, c}(x,\theta,\bar{\theta})$ is only defined on $\mathcal{M}$. 

We now show that $\Phi_{\mathcal{M}\,a}$ has the correct transformation property of an undotted spinor with the label $\left(\frac{1}{2},0,q,\bar{q}\right)$. Eq.~(\ref{UndottedProjection}) gives
\begin{equation}
\delta_s \Phi_{\mathcal{M}\,a} = \left(X^+\right)^{q-\frac{1}{2}} \left(\bar{X}^+\right)^{\bar{q}} \delta_s \Phi_{\mathcal{E}\,A=a} + \left(q-\frac{1}{2}\right)\left(\frac{\delta X^+}{X^+}\right)\Phi_{\mathcal{M}\,a} + \bar{q}\left(\frac{\delta \bar{X}^+}{\bar{X}^+}\right) \Phi_{\mathcal{M}\,a},
\end{equation}  
where the variation $\delta_s$ on a field $\Phi(X)$ is defined by $\delta_s \Phi(X) = \Phi'(X'(X))-\Phi(X)$ (in other words, we omit the action of $SU(2,2|1)$ acting on the coordinates in Eq.~(\ref{XParametrize})).    Parametrizing the $SU(2,2|1)$ generator $T_A^{\phantom{B}B}$ in Eq.~(\ref{TAB}) in terms of the more familiar four-dimensional superconformal transformations:  the translation $a_\mu$, $SO(3,1)$ transformation $\omega_{\mu\nu}$, dilatation $\lambda$, special conformal transformation $b_\mu$, Poincare supersymmetry  $\tau_a$, special superconformal transformation $\eta_a$, and $U(1)_R$ charge $\phi$,
\begin{equation}
T_A^{\phantom{A}B} = \left(
\begin{array}{ccc}
\frac{i}{2}\mathrm{log}\lambda\delta_a^{\phantom{a}b} + \frac{i}{2}\omega_{\mu\nu}\left(\sigma^{\mu\nu}\right)_a^{\phantom{a}b} + \frac{1}{4}\phi\delta_a^{\phantom{a}b} & b_\mu \sigma^\mu_{a\dot{b}} & -2\eta_a \\
a_\mu\bar{\sigma}^{\mu\dot{a}b} & -\frac{i}{2}\mathrm{log}\lambda\delta^{\dot{a}}_{\phantom{a}\dot{b}} + \frac{i}{2}\omega_{\mu\nu}\left(\bar{\sigma}^{\mu\nu}\right)^{\dot{a}}_{\phantom{a}\dot{b}} + \frac{1}{4}\phi\delta^{\dot{a}}_{\phantom{a}\dot{b}} & 2\bar{\tau}^{\dot{a}} \\
2\tau^b & -2\bar{\eta}_{\dot{b}} & \phi \\
\end{array} \right),
\label{TABParametrize}
\end{equation}
we obtain 
\begin{eqnarray} 
\delta X^+ = \left[-\mathrm{log}\lambda + \frac{i}{2}\phi - 4\eta^a\theta_a + 2(b\cdot y)\right] X^+, 
\label{deltaX+} \\
\delta \bar{X}^+ = \left[-\mathrm{log}\lambda - \frac{i}{2}\phi - 4\bar{\eta}_{\dot{a}}\bar{\theta}^{\dot{a}} + 2(b\cdot \bar{y})\right] \bar{X}^+,
\label{deltaXbar+}
\end{eqnarray}
and, using  Eq.~(\ref{UndottedConstraintComponents}),
\begin{equation}
\delta_s \Phi_{\mathcal{E}\,a} = -\frac{1}{2}\omega_{\mu\nu} \left(\sigma^{\mu\nu}\right)_a^{\phantom{a}b} \Phi_{\mathcal{E}\,b} - \frac{1}{2}\mathrm{log}\lambda \Phi_{\mathcal{E}\,a} + \left(b\cdot y\right)\Phi_{\mathcal{E}\,a} - 2b_\mu y_\nu \left(\sigma^{\mu\nu}\right)_a^{\phantom{a}b} \Phi_{\mathcal{E}\,b} + \frac{i}{4}\phi \Phi_{\mathcal{E}\,a} + 4\eta_a\theta^b \Phi_{\mathcal{E}\,b}.
\label{PhiaTransfConstraint}   
\end{equation}
Note that this field has zero variation under translations and Poincare supersymmetry transformations.    Putting together these results, we finally get 
\begin{eqnarray}
\delta_s \Phi_{\mathcal{M}\,a}= && -\frac{1}{2}\omega_{\mu\nu} \left(\sigma^{\mu\nu}\right)_a^{\phantom{a}b} \Phi_{\mathcal{M}\,b} -\mathrm{log}\lambda\left(q+\bar{q}\right) \Phi_{\mathcal{M}\,a} + 2q\left(b\cdot y\right)\Phi_{\mathcal{M}\,a} + 2\bar{q}\left(b\cdot\bar{y}\right)\Phi_{\mathcal{M}\,a} \nonumber \\
&& - 2b_\mu y_\nu \left(\sigma^{\mu\nu}\right)_a^{\phantom{a}b} \Phi_{\mathcal{M}\,b} - 4\left(q-\frac{1}{2}\right) (\eta\theta) \Phi_{\mathcal{M}\,a} + 4\eta_a\theta^b \Phi_{\mathcal{M}\,b} - 4\bar{q}\left(\bar{\eta}\bar{\theta}\right)\Phi_{\mathcal{M}\,a} \nonumber \\
&& + \frac{i}{2}\phi\left(q-\bar{q}\right)\Phi_{\mathcal{M}\,a},
\end{eqnarray}
which is the correct transformation rule for a $(1/2,0)$ spinor multiplet, see e.g., ref.~\cite{Buchbinder:1998qv}. The parameter $\phi$ does not correspond to the standard $R$-symmetry charge assignment in four dimensions, but differs by a factor of $\frac{4}{3}$, see Ref.~\cite{GSS}.

To summarize:   A field $\Phi_{\mathcal{E}\,A}(X,{\bar X})$ on ${\cal E}$ obeying the scaling relation (i) $\Phi_{\mathcal{E}\,A}\left(\lambda X,\bar{\lambda} \bar{X}\right) = \lambda^{-\left(q-\frac{1}{2}\right)} \bar{\lambda}^{-\bar{q}} \Phi_{\mathcal{E}\,A}\left(X,\bar{X}\right)$ and the covariant constraint (ii)  $[X_{AB} \Phi_{{\cal E}\,C}]_{\bf 15}=0$ defines a superconformal multiplet $\Phi_{\mathcal{M}\,a}(x,\theta,{\bar\theta})$ in the $(1/2,0,q,{\bar q})$ representation via the relation in Eq.~(\ref{UndottedProjection}).   Conversely, given $\Phi_{\mathcal{M}\,a}(x,\theta,{\bar\theta})$ we can construct an operator  $\Phi_{\mathcal{E}\,A}\left(X, \bar{X}\right)$ that transforms linearly under $SU(2,2|1)$.

This construction generalizes easily to other representations $(j,{\bar j},q,{\bar q})$.    A superfield $\Phi_{\mathcal{M}\,\dot{a}}$ transforming in the $\left(0,\frac{1}{2}, q,\bar{q}\right)$ representation corresponds in  $\mathcal{E}$ to a superfield $\Phi_{\mathcal{E}}^A$ that obeys the scaling property
\begin{equation}
\Phi_{\mathcal{E}}^A \left(\lambda X,\bar{\lambda}\bar{X}\right) = \lambda^{-q}\bar{\lambda}^{-\left(\bar{q}-\frac{1}{2}\right)} \Phi_{\mathcal{E}}^A \left(X,\bar{X}\right), \hspace{5mm} \mathrm{and}
\label{DottedHomog}
\end{equation}
and the constraint $[\Phi_{\mathcal{E}}^A {\bar X}^{BC}]_{\bf {\overline{15}}}=0$, or 
\begin{equation}
 \Phi_{\mathcal{E}}^{\underline{A}} \bar{X}^{\underline{B}\underline{C}} = 0,
\label{DottedConstraint}
\end{equation}
which implies the addition constraint $\Phi_{\mathcal{E}}^{A}X_{AB}=0$ on $\mathcal{M}$. One obtains $\Phi_{\mathcal{M}\,\dot{a}}$ from $\Phi_{\mathcal{E}}^{A}$ by 
\begin{equation}
\Phi_{\mathcal{M}\,\dot{a}} \left(x,\theta,\bar{\theta}\right) = \left(X^+\right)^{q} \left(\bar{X}^+\right)^{\bar{q}-\frac{1}{2}} \Phi_{\mathcal{E}}^{A=\dot{a}} \left(X,\bar{X}\right),
\label{DottedProjection} 
\end{equation}
and conversely
\begin{equation}
  \Phi^{A}_{\mathcal{E}}(X,\bar{X})=\left(X^{+}\right)^{-q}\left(\bar{X}^{+}\right)^{-(\bar{q}+\frac{1}{2})}\Phi_{\mathcal{M}\,\dot{c}}(x,\theta,\bar{\theta})\bar{X}^{\dot{c}A}
  \label{DottedInverseProjection} 
\end{equation}
defines a field $\Phi^{A}_{\mathcal{E}}(X,\bar{X})$ that transforms linearly under superconformal transformations.

For the more general case of  $\Phi_{\mathcal{M}\,a_1\cdots a_{2j}}{}^{\dot{b}_1\cdots \dot{b}_{2{\bar j}}} \left(x,\theta,\bar{\theta}\right)$ in the $(j,{\bar j},q,{\bar q})$ representation, we introduce a field 
$\Phi_{\mathcal{E}\,A_1\cdots A_{2j}} {}^{B_1\cdots B_{2{\bar j}}}(X,{\bar X})$, which we assign to have identical $SU(2,2|1)$ transformation property to that of the tensor product of $2j$ fundamentals $V_{A_1}^{(1)} \ldots  V_{A_{2j}}^{(2j)}$ and $2 \bar{j}$ anti-fundamentals $\bar{V}^{B_1}_{(1)} \ldots  \bar{V}^{B_{2\bar{j}}}_{(2\bar{j})}$
\begin{equation}
\Phi_{\mathcal{E}\,A_1\cdots A_{2j}}{}^{B_1\cdots B_{2{\bar j}}}\sim V_{A_1}^{(1)}\cdots V_{A_{2j}}^{(2j)} \, {\bar V}^{B_1}_{(1)} \cdots {\bar V}^{B_{2{\bar j}}}_{(2\bar{j})}.
\label{eq:trandsym}
\end{equation}
We pick the highest-weight $(j,\bar{j})$ representation by imposing a symmetry  under the exchange of adjacent indices $\Phi_{\mathcal{E}\,\cdots A_i A_{i+1} \cdots}{}^{B_1\cdots B_{2{\bar j}}}=
   - \sigma(A_i A_{i+1}) \Phi_{\mathcal{E}\,\cdots A_{i+1} A_i  \cdots}{}^{B_1\cdots B_{2{\bar j}}}$.
The fields are homogenous 
\begin{equation}
\Phi_{\mathcal{E}\,A_1\cdots A_{2j}}^{\phantom{\mathcal{E}\,A_1\cdots A_{2j}}B_1\cdots B_{2\bar j}}\left(\lambda X,\bar{\lambda}\bar{X}\right) = \lambda^{-\left(q-j\right)} \bar{\lambda}^{-\left(\bar{q}-{\bar j}\right)} \Phi_{\mathcal{E}\,A_1\cdots A_{2j}}^{\phantom{\mathcal{E}\,A_1\cdots A_{2j}}B_1\cdots B_{2\bar j}}\left(X,\bar{X}\right)
\label{PhimnHomog}
\end{equation}
and satisfy the constraints
\begin{eqnarray}
\label{eq:multcon1}
&&0 = X_{\underline{C}\underline{D}} \Phi_{\mathcal{E}\,\underline{A}_1A_2\cdots A_{2j}}^{\phantom{\mathcal{E}\,\underline{A}_1A_2\cdots A_{2j}}B_1\cdots B_{2\bar j}}, \hspace{10mm} \\
\label{eq:multcon2}
&&0 = \Phi_{\mathcal{E}\,A_1\cdots A_{2j}}^{\phantom{\mathcal{E}\,A_1\cdots A_{2j}}B_1\cdots B_{2\bar j-1}\underline{B}_{2\bar j}} \bar{X}^{\underline{C}\underline{D}}.
\end{eqnarray}
Using the quadratic equations satisfied by points $(X_{AB},{\bar X}^{AB})$ on ${\cal M}$, these constraints also imply that
\begin{eqnarray}
\label{eq:xbarcon}
{\bar X}^{AA_1}\Phi_{\mathcal{E}\, A_1A_2\cdots A_{2j}}^{\phantom{\mathcal{E}\,\underline{A}_1A_2\cdots A_{2j}}B_1\cdots B_{2\bar j}}=0,\\
\label{eq:xcon}
\Phi_{\mathcal{E}\, A_1A_2\cdots A_{2j}}^{\phantom{\mathcal{E}\,\underline{A}_1A_2\cdots A_{2j}}B_1\cdots B_{2\bar j}} X_{B_{2\bar j}A}=0.
\end{eqnarray}
The superfield $\Phi_{\mathcal{M}\,a_1\cdots a_{2j}}{}^{\dot{b}_1\cdots \dot{b}_{2\bar j}}$ is recovered through
\begin{equation}
\Phi_{\mathcal{M}\,a_1\cdots a_{2j}}{}^{\dot{b}_1\cdots \dot{b}_{2\bar j}} \left(x,\theta,\bar{\theta}\right) = \left(X^+\right)^{q-j}\left(\bar{X}^+\right)^{\bar{q}-{\bar j}} \Phi_{\mathcal{E}\,A_1=a_1\cdots A_{2 j}=a_{2 j}}{}^{B_1=\dot{b}_1\cdots B_{2 {\bar j}}=\dot{b}_{2{\bar j}}}
\label{PhimnProjection}
\end{equation}
and one can also easily generalize Eq.~(\ref{UndottedInverseProjection}) to this case. 

It is sometimes possible to further reduce the superfield by imposing holomorphy or anti-holomorphy, i.e.  functional dependence only on  ${ X}_{AB}$ or ${\bar X}_{AB}$ respectively.   Given the constraints in Eqs.~(\ref{eq:multcon1}), (\ref{eq:multcon2}), holomorphic superfields $\Phi_{\cal E}(X)$ on ${\cal E}$ project onto chiral superfields $\Phi_{\cal M}(y,\theta)$ only for the special values ${\bar j}={\bar q}=0$ of the quantum numbers.   In particular,  this implies the standard relation $\Delta=\frac{3}{2} R$ between scaling dimension and $R$-charge in the chiral sector (in a normalization with $R(\theta_a)=1$).   Likewise, anti-chiral fields $\Phi_{\cal M}({\bar y},{\bar \theta})$ on ${\cal M}$ correspond to anti-holomorphic fields on ${\cal E}$ with $j=q=0$, and consquently $\Delta=-\frac{3}{2}  R$.   

In what follows, we will focus on special cases of $(j,{\bar j},q,{\bar q})$ that have particular relevance to physical applications.  In addition to chiral/anti-chiral fields with $j={\bar j}=0$ we will consider the real scalar multiplet $V_{\cal M}$ transforming in the representation $(0,0,q,q)$.   The case $q=1$, i.e. $\Delta=2$, $R=0$, usually denoted by $L_{\cal M}$, contains a dimension $\Delta=3$ conserved current $j^\mu(x)$.    This multiplet can be obtained from a real multiplet $L_{\cal M}$ by imposing the constraints~\cite{Osborn:1998qu} $D^2 L_{\mathcal{M}} = \bar{D}^2 L_{\mathcal{M}} = 0$, where $D_a, {\bar D}^{\dot a}$ are the ${\cal N}=1$ Poincare super-covariant derivatives, which restrict  $L_{\mathcal{M}}$ to
\begin{eqnarray}
\nonumber
L_{\mathcal{M}}\left(x,\theta,\bar{\theta}\right) & = & C(x) + i\theta\chi(x) - i\bar{\theta}\bar{\chi}(x)  - \theta\sigma^\mu\bar{\theta}j_\mu(x) + \frac{1}{2} \theta^2\bar{\theta}\bar{\sigma}^\mu\partial_\mu\chi(x)-\frac{1}{2}  \bar{\theta}^2\theta\sigma^\mu\partial_\mu\bar{\chi}(x)\\
 & &  - \frac{1}{4} \theta^2\bar{\theta}^2 \Box C(x).
\label{VMComponents}
\end{eqnarray}

Finally, we will also consider the supercurrent multiplet~\cite{Ferrara:1974pz}, a real superfield ${{\cal T}_{\cal M}}_{a\dot b}(x,\theta,{\bar\theta})$ transforming in the $(\frac{1}{2},\frac{1}{2},\frac{3}{2},\frac{3}{2})$ representation and obeying the conservation law $D^a {{\cal T}_{\cal M}}_{a\dot b} = {\bar D}^{\dot b} {{\cal T}_{\cal M}}_{a\dot b}=0$ with component field expansion
\begin{equation}
\mathcal{T}_{\mathcal{M}\,\mu}\left(x,\theta,\bar{\theta}\right) = j_\mu^R(x) + \theta^a S_{\mu\,a}(x) + \bar{\theta}_{\dot{a}} \bar{S}_\mu^{\dot{a}}(x) + 2\theta\sigma^\nu\bar{\theta} T_{\nu\mu}(x) + \dots,  
\end{equation}
where $j_\mu^R$ is the $U(1)_R$ current, $S_{\mu\, a }$ the super-current, and $T_{\mu\nu}=T_{\nu\mu}$ the energy-momentum tensor.   In the embedding approach,  ${{\cal T}_{\cal M}}_{a\dot b}$  gets lifted to a superfield ${{\cal T}_{\cal E}}_A{}^B$ in the ${\bf 24}$ (adjoint) representation, satisfying the constraints of  Eqs.~(\ref{eq:multcon1}), (\ref{eq:multcon2}), ${\rm str} \, {\cal T}=0$,  together with the scaling law 
\begin{equation}
\mathcal{T}_{\mathcal{E}\,A}^{\phantom{EA}B}\left(\lambda X,\bar{\lambda} \bar{X}\right) = \lambda^{-1}\bar{\lambda}^{-1}  \mathcal{T}_{\mathcal{E}\,A}^{\phantom{EA}B}\left(X,\bar{X}\right),
\label{THomog}
\end{equation}
and the reality condition
\begin{equation}
\mathcal{T}_{\mathcal{E}\,A}^{\phantom{EA}B}(X,{\bar X}) = \bar{\mathcal{T}}_{\mathcal{E}\, A}^{\phantom{E A}B}(X,{\bar X}) \equiv A_{A\dot{B}}  \mathcal{T}_{\mathcal{E}\,\phantom{\, A}\dot{A}}^{\dagger \, \dot{B}}  (X,{\bar X})A^{\dot{A}B}.
\label{TReality}
\end{equation}
The relation between ${\cal T}_{\cal M}$ and ${\cal T}_{\cal E}$ is 
\begin{equation}
\mathcal{T}_{\mathcal{M}\,a\dot{b}}\left(x,\theta,\bar{\theta}\right) = X^+ \bar{X}^+ \mathcal{T}_{\mathcal{E}\,A=a}{}^{B={\dot b}}\left(X,\bar{X}\right),
\label{TProjection}
\end{equation}
and
\begin{equation}
 \mathcal{T}_{\cal{E}\,A}^{\phantom{A}B}(X,\bar{X})=\left(X^{+}\right)^{-2}\left(\bar{X}^{+}\right)^{-2}X_{A}^{\phantom{A}a}\, \mathcal{T}_{a\dot{b}}(x,\theta,{\bar\theta})
 \, \bar{X}^{\dot{b}B}.
  \label{TInverseProjection}
\end{equation}

\section{Correlators}
\label{Correlators}
We now use the superembedding formalism to construct manifestly covariant expressions for SCFT correlation functions.

\subsection{2-point functions}
Given a set of coordinates $Z_i=(X_{i},{\bar X}_{i})$ an over-complete set of  $SU(2,2|1)$ invariants is given by the supertraces~\cite{Jarvis:1978bc}, e.g.
\begin{equation}
\langle 1 {\bar 2} 3 {\bar 4} \cdots \rangle \equiv \mbox{str} X_1 \sigma {\bar X}_2 X_3\cdots,
\end{equation}
where the rules for constructing the tensor $(X_1 \sigma {\bar X}_2 X_3\cdots)_{A}{}^B$ were given in sec.~\ref{Superembedding Formalism}.   For two independent  points, the only invariant is $\langle 1{\bar 2}\rangle$ and its complex conjugate, as can be readily seen by going to the frame in which $X_1$ is at the origin and $X_2$ at infinity.    It follows from the scaling relation Eq.~(\ref{PhimnHomog}) that, up to  normalization,
\begin{equation}
\left\langle \Phi_{\mathcal{E}1}\left(Z_1\right) \Phi_{\mathcal{E}2}\left(Z_2 \right) \right\rangle = \frac{\delta_{q_1\bar{q}_2} \delta_{\bar{q}_1q_2}}{\langle 1\bar{2}\rangle^{q_1} \langle 2\bar{1}\rangle^{\bar{q}_1}}.
\label{Scalar2PF}
\end{equation}  
The four-dimensional correlator immediately follows upon inserting the expression
\begin{equation}
\left\langle 1\bar{2}\right\rangle = -\frac{1}{2} \left(X_1^+\right) \left(\bar{X}_2^+\right) \left({\bar y}_2 - y_1 +2i\theta_1\sigma\bar{\theta}_2\right)^2.
\label{ijbar}
\end{equation}
Eq.~(\ref{Scalar2PF}) contains as a special case the two-point function of a chiral (${\bar q}=0$) scalar with an antichiral ($q=0$) scalar, and the two-point function of the current superfield $L_{\cal E}\sim (0,0,1,1)$.   In the normalization of~\cite{Fortin:2011nq}, with $\tau$ a real constant
\begin{equation}
\left\langle L^I_{\mathcal{E}}\left(Z_1\right) L^J_{\mathcal{E}}\left(Z_2\right) \right\rangle =\frac{\delta^{IJ}}{64 \pi^4} \frac{\tau}{\left\langle 1\bar{2}\right\rangle \left\langle 2\bar{1}\right\rangle}\, ,
\label{L2PF}
\end{equation}  
where $I,J$ are the adjoint indices of the symmetry group.  This result holds up to contact terms, which the embedding formalism does not account for~\cite{GSS}.  It is straightforward to check that upon projection to ${\cal M}$, the correlator satisfies $D^2 \left\langle L_{\mathcal{E}} L_{\mathcal{E}} \right\rangle = {\bar D}^2\left\langle L_{\mathcal{E}} L_{\mathcal{E}} \right\rangle=0$.  Thus, current conservation is automatic at the level of the two-point function.

To construct correlators for higher-spin supermultiplets, we employ a variant of the index-free notation introduced in~\cite{Costa:2011mg} for $SO(4,2)$ tensors and in~\cite{SimmonsDuffin:2012uy} for $SU(2,2)$ multi-twistors.   We use fundamental and anti-fundamental representations $W_A$, ${\bar W}^A$ whose components   are now $W_\alpha$ and ${\bar W}^\alpha$  $c$-numbers, while components $W_5,{\bar W}^5$ are Grassmann variables.    We write
\begin{equation}
\Phi(W,{\bar W},Z) = \sum_{j,{\bar j}}\sigma(\cdots) {\bar W}^{A_1}\cdots {\bar W}^{A_{2j}} \Phi_{{\cal E}A_1\cdots A_{2j}}{}^{B_1\cdots B_{2\bar j}} (Z) W_{B_1}\cdots W_{B_{2\bar j}}.
\end{equation}
which is a superconformal scalar under simultaneous transformations of ${\Phi}_{{\cal E}A_1\cdots A_{2j}}{}^{B_1\cdots B_{2{\bar j}}}$ , $X_{AB}$, ${\bar X}^{AB}$,  $W_A$, ${\bar W}^A$.  (The symbol $\sigma(\cdots)$ denotes factors of Eq.~(\ref{eq:sigma}) inserted to ensure $SU(2,2|1)$ invariance).

Correlators of $\Phi(W,{\bar W},Z)$ are functions of invariants constructed from insertions of the objects $W$, ${\bar W}$, $X$ and ${\bar X}$.   Note that due to the constraints in Eqs.~(\ref{eq:xbarcon}),~(\ref{eq:xcon}),  there is an additional ``gauge invariance'' under the shifts 
\begin{eqnarray}
W_A            &\rightarrow& W_A      +  {\bar S}^B X_{BA}, \label{eq:gaugeW}\\
{\bar W}^A &\rightarrow& {\bar W}^A + {\bar X}^{AB} S_B ,\label{eq:gaugeWbar}
\end{eqnarray}
with arbitrary $S_A, {\bar S}^A$, which can be used to reduce the number of invariants.   For the two-point function $\langle\Phi_1 \Phi_2\rangle$, a complete set of invariants on ${\cal M}$ consists of $\langle 1 {\bar 2}\rangle$ and its conjugate, together with
\begin{eqnarray}
\label{eq:ginv}
{\bar W}_1{} (1 {\bar 2}) W_2{},\\
\label{eq:ginvc}
{\bar W}_2{}  (2  {\bar 1}){} W_1{},
\end{eqnarray}
where ${\bar W}_1{} (1 {\bar 2}) W_2{} \equiv {\bar W}_1{} (X_1\sigma {\bar X}_2)  W_2{}$, etc. In the following discussion, we will abbreviate $X_1$ as $1$,  ${\bar X}_1$ as $\bar 1$ and so on and omit the $\sigma$'s as their position is uniquely specified by $SU(2,2|1)$ invariance.  The gauge transformations on $W_A,{\bar W}^A$ introduced above forbid $SU(2,2|1)$ invariants like ${\bar W}_1{} (2 {\bar 1})  W_2$ and others, while the exchange symmetries ${\bar W}^A {\bar W}^B=-\sigma(AB) \sigma(A) \sigma(B) {\bar W}^B {\bar W}^A$ and ${X}_{AB}=\sigma(AB) {X}_{BA}$ rule out invariants such as ${\bar W}_1 (1)  {\bar W}_1=0$. Longer strings of coordinates either vanish due to the $[X \bar{X}]_{\bf{24}}=0$ constraint, or can be reduced to the basic invariants using the $[X X]_{\bf{16}}=0$ constraint. For example,  $(1\bar{2}1\bar{2})_A^{\phantom{A}B}$ is proportional to $\langle 1 \bar{2}\rangle$ times $(1\bar{2})_A^{\phantom{A}B}$. Finally, by the scaling properties all two-point functions reduce to powers of ${\bar W}_1{} (1 {\bar 2}) W_2{}$, ${\bar W}_2{}  (2  {\bar 1}){} W_1{}$  and $\langle 1 {\bar 2}\rangle$, $\langle 2 {\bar 1}\rangle$.

We find using these results that the two point function of  $\Phi_1\sim (j,{\bar j},q,{\bar q})$ with another superfield $\Phi_2$ is non-vanishing only for $\Phi_2$ transforming in the representation $({\bar j},j,{\bar q},q)$.  This two-point funtion can be read off the term in $\langle \Phi_1 \Phi_2 \rangle$ proportional to $2j$ powers of ${W_2}_A, {\bar W}_1^B$ and $2{\bar j}$ powers of  ${W_1}_A, {\bar W}_2^B$.   It is
\begin{equation}
\frac{\left[ {\bar W}_1 (1 {\bar 2}) W_2\right]^{2j} \left[{\bar W}_2{} (2  {\bar 1}) W_1{}\right]^{2\bar j}}{\langle 1{\bar 2}\rangle^{q+j}  \langle 2 {\bar 1}\rangle^{{\bar q}+{\bar j}}}.
\end{equation}
For example, the two-point function of superfields ${\Phi_1}_A(Z)\sim (\frac{1}{2},0,q,{\bar q})$ and ${\Phi_2}^A(Z)\sim (0,\frac{1}{2},{\bar q}, q)$ has the form $\langle {\Phi_1}_A  {\Phi_2}^B\rangle  =  \langle 1{\bar 2}\rangle^{-q-\frac{1}{2}}\langle 2 {\bar 1\rangle^{-{\bar q}}}(1\bar{2})_A{}^B$.  Using
\begin{equation}
\left(1\bar{2}\right)_{A=a}{}^{B={\dot b}} = -\frac{i}{4} X^+_1 {\bar X}^+_2 (y_{1{\bar 2}}+4 i \theta_1{\bar\theta}_2)_{a{\dot b}}
\end{equation}   
($y_{{\bar 1}2}\equiv {\bar y}_1 - y_2=-y_{2{\bar 1}}$), together with the rules given in Sec.~\ref{Dictionary} for projecting onto four-dimensional superfields then yields 
\begin{equation}
\langle {\phi_1}_{a}(x_{1},\theta_{1},{\bar\theta}_1 ){\phi_2}_{\dot{b}}(x_{2},\theta_2,\bar{\theta}_{2})\rangle=\frac{\left(y_{1\bar{2}}+4i\theta_{1}\bar{\theta}_{2}\right)_{a\dot{b}}     }{     \left[\left(y_{\bar{2}1}+2i\theta_1\sigma\bar{\theta}_2\right)^2\right]^{q+1/2}\left[\left(y_{\bar{1}2}+2i\theta_2\sigma\bar{\theta}_1\right)^2\right]^{{\bar q}}} \,  .
\end{equation}
Similarly the two-point function of the supercurrent $\mathcal{T}_{a\dot{b}}(x,\theta,\bar{\theta})\sim(\frac{1}{2},\frac{1}{2},\frac{3}{2},\frac{3}{2})$ is given by the embedding space expression
\begin{equation}
   \langle \left[{\bar W}_1{} \mathcal{T} (Z_1)W_1 \right]\left[ {\bar W}_2 \mathcal{T}_{C}(Z_2) {W_{2}}\right]\rangle = \frac{c_{\mathcal{TT}} }{  16} \frac{\left[ {\bar W}_1 (1\bar{2}) W_2\right] \left[{\bar W}_2{} (2 \bar{1}) W_1{}\right]    }{\langle 1{\bar 2}\rangle^{2}  \langle 2 {\bar 1}\rangle^2},
 \end{equation}
or in four-dimensional language
\begin{equation}
\left\langle T_{a\dot{b}} \left(x_1,\theta_1,\bar{\theta}_1 \right) T_{c\dot{d}} \left(x_2,\theta_2,\bar{\theta}_2 \right) \right\rangle = -\frac{c_{\mathcal{TT}}   }{ 16}
 \frac{\left( y_{1\bar{2}} + 4i\theta_1\bar{\theta}_2 \right)_{a\dot{d}} \left( y_{2\bar{1}} + 4i\theta_2\bar{\theta}_1 \right)_{c\dot{b}}}{\left[\left(y_{\bar{2}1}+2i\theta_1\sigma\bar{\theta}_2\right)^2\left(y_{\bar{1}2}+2i\theta_2\sigma\bar{\theta}_1\right)^2\right]^2},
\end{equation}
in agreement with Ref.~\cite{Osborn:1998qu}.

\subsection{3-point functions}
Supertraces constructed from strings of products of the three coordinates $(X,{\bar X})_{i=1,2,3}$ reduce to products of the bilinears $\langle i {\bar j}\rangle$, as can be seen for instance in a frame in which two points are fixed to the origin and infinity respectively.    The remaining symmetries in this frame are sufficient to fix the bosonic part of the third point, but not to set its Grassmann part to zero.   Thus, in contrast to  non-supersymmetric theories, there is an invariant cross ratio for three points~\cite{Osborn:1998qu}.  It can be taken to be
 \begin{equation}
u = \frac{\left\langle 1\bar{2}\right\rangle \left\langle 2\bar{3}\right\rangle \left\langle 3\bar{1}\right\rangle }{\left\langle 2\bar{1}\right\rangle \left\langle 3\bar{2}\right\rangle \left\langle 1\bar{3}\right\rangle}.
\end{equation}

Because superfield three-point correlators can have arbitrary dependence on $u$, it might seem at first that predictive power is completely lost.   Fortunately, the functional dependence on $u$ is fixed up to three numerical constants:  In the frame with points $X_{1,2}$ fixed to the origin and infinity respectively, there is residual  $SL(2,{\mathbb C})\times U(1)_R$ plus dilation symmetry that fixes the unique independent invariant to be  $\theta_3 x_3\cdot \sigma{\bar\theta}_3/x_3^2$.   This invariant  is related to $u$ by
\begin{equation}
  z \equiv     \frac{1-u  }{ 1+u}=-\frac{2i\theta_3 x_3\cdot \sigma{\bar\theta}_3  }{ x_3^2} .
\end{equation}    
Thus, in any frame, the most general function $f(u)$ is a quadratic polynomial in $z$ so  $SU(2,2|1)$ symmetry does yield some predictions, in the form of relations between component field correlators.    Many known 4D SCFTs have in addition global symmetries, whose (possibly anomalous) Ward identities provide extra constraints.

Consider, for instance, the three-point function of conserved currents $L^I(x,\theta,{\bar\theta})$.   By including suitable improvement terms if necessary, the correlator can be made symmetric in the exchange of operator labels.    Thus in terms of the structure constants $f^{IJK}$ and anomaly tensor $d^{IJK}$ of the global symmetry group $G$, 
\begin{equation}
\left\langle L^I_{\mathcal{E}} (Z_1) L^J _{\mathcal{E}}(Z_2) L_{\mathcal{E}}^K(Z_3) \right\rangle = \frac{d^{IJK}\left(\lambda_0 + \lambda_2 z^2\right) + \lambda_1 z \, f^{IJK}}{\left[\left\langle 1\bar{2}\right\rangle\left\langle 2\bar{1}\right\rangle\left\langle 1\bar{3}\right\rangle\left\langle 3\bar{1}\right\rangle\left\langle 2\bar{3}\right\rangle\left\langle 3\bar{2}\right\rangle\right]^{\frac{1}{2}}}
\label{L3PFNotQuite}
\end{equation}
We have used the property $z\rightarrow -z$ under interchange of coordinates to fix the dependence on $u$.

To fix the constants $\lambda_{0,1,2}$ we impose Ward identites that relate the coefficient of $f^{IJK}$ to the two-point function and the coefficient of $d^{IJK}$ to the $\mbox{Tr}\, G^3$ chiral anomaly.   Because we lack an embedding space formulation of contact terms, we impose these Ward identities directly on the component fields.   First send $x_1\rightarrow 0$ and $x_2\rightarrow\infty$, and pick out the coefficient of $\theta_3{\bar\theta}_3$ from Eq.~(\ref{L3PFNotQuite}):
\begin{equation}
\langle C^I(x_1\rightarrow 0) C^J(x_2\rightarrow\infty) j^K_\mu(x_3) \rangle = - {8i\lambda_1 f^{IJK} \over x^4_2} {\partial\over\partial x^\mu_3} \left({x^2_3}\right)^{-1},
\end{equation}
where $C^I(x)=L^I(x,\theta={\bar\theta}=0)$.  From the $\langle C^I C^J\rangle$ component of Eq.~(\ref{L2PF}) and the $G$-symmetry Ward identity
\begin{equation}
\partial_3^\mu \langle TC^I(x_1) C^J(x_2) j^K_\mu(x_3) \rangle = -if^{IJK}\left[\delta^4(x_1-x_3) - \delta^4(x_2-x_3)\right]  \frac{\tau}{16\pi^4 x^4_{12}},
\end{equation}
we find, using $\Box (x^2)^{-1} =4\pi^2i\delta^4(x)$, that $\lambda_1=- i\tau/(512\pi^6)$.  To fix $\lambda_2$ in terms of $\lambda_0$ note that  $\langle C^I C^J C^K\rangle$  is
\begin{equation}
\langle C^I(x_1\rightarrow 0) C^J(x_2\rightarrow\infty) C^K(x_3) \rangle = {8\over x_2^4 x_3^2} d^{IJK} \, \lambda_0,
\end{equation}
while the $\theta^2_3 {\bar\theta}^2_3$ component gives
\begin{equation}
\langle C^I(x_1\rightarrow 0) C^J(x_2\rightarrow\infty) \Box C^K(x_3) \rangle =  -{32\over x_2^4 x_3^4} d^{IJK}\left(2\lambda_2-\lambda_0\right),
\end{equation}
where $L(x,\theta,{\bar\theta})|_{\theta^2{\bar\theta}^2}=-{1\over 4}\Box C(x)$.  Up to contact terms, these two results are only consistent if $\lambda_2={1\over 2}\lambda_0$.  Finally, $\lambda_0$ is fixed in terms of the chiral anomaly, which from Eq.~(\ref{L3PFNotQuite}) is given by
\begin{eqnarray}
\nonumber
\langle j_{\mu_1}^I(x_1) j_{\mu_2}^J(x_2) j_{\mu_3} ^K(x_3) \rangle|_d &=&  {8 i(\lambda_0+6\lambda_2)\over x_{12}^4  x_{23}^4 x_{31}^4} d^{IJK} \Big(\mbox{Tr}\left[\sigma\cdot x_{12} {\bar\sigma}_{\mu_2} \sigma\cdot x_{23} {\bar\sigma}_{\mu_3} \sigma\cdot x_{31} {\bar\sigma}_{\mu_1}\right] \\
& & \,\,\,\,-  {}\ \mbox{Tr}\left[\sigma\cdot x_{31} {\bar\sigma}_{\mu_3} \sigma\cdot x_{23} {\bar\sigma}_{\mu_2} \sigma\cdot x_{12} {\bar\sigma}_{\mu_1}\right]\Big),
\end{eqnarray}
which in turn is equal to the one-loop anomaly of $k$ free chiral fermions, provided we adjust $\lambda_0=k/(1024\pi^6)$.   Thus, the three-point function can be written as
\begin{equation}
\left\langle L^I_{\mathcal{E}} (Z_1) L^J _{\mathcal{E}}(Z_2) L_{\mathcal{E}}^K(Z_3) \right\rangle ={1\over 1024\pi^6} \frac{ - 2 i\tau f^{IJK}z +k d^{IJK}\left(1 + {1\over 2} z^2\right)}{\left[\left\langle 1\bar{2}\right\rangle\left\langle 2\bar{1}\right\rangle\left\langle 1\bar{3}\right\rangle\left\langle 3\bar{1}\right\rangle\left\langle 2\bar{3}\right\rangle\left\langle 3\bar{2}\right\rangle\right]^{\frac{1}{2}}}.
\end{equation}

The correlator $\langle {\mathcal T} L^I L^J\rangle$ is obtained by similar considerations.   Imposing $SU(2,2|1)$ invariance, symmetry under exchange $(1,I)\leftrightarrow (2,J)$, and reality conditions on the fields, the most general form is 
\begin{eqnarray}
\nonumber
 \hspace{-20pt} \langle L^I(Z_1) L^J (Z_2) \left[{\bar W}_3 \mathcal{T} (Z_3)W_3 \right]  \rangle &=&  {\delta^{IJ} \over\langle 3 {\bar 1} \rangle \langle1 {\bar 3}\rangle   \langle 3 {\bar 2} \rangle \langle 2 {\bar 3}\rangle}  \left[
 {\hat\lambda}_0 \cdot {\bar W}_3\left(\frac{3 {\bar 2} 1{\bar 3}}{\langle 1 {\bar 2}\rangle} + \frac{3 {\bar 1} 2 {\bar 3}}{\langle 2 {\bar 1}\rangle}\right) W_3  \right.\\
 & & \left. +\,  z  {\hat\lambda}_1 \cdot  {\bar W}_3\left( \frac{3 {\bar 2} 1{\bar 3}}{\langle 1 {\bar 2}\rangle} - \frac{3 {\bar 1} 2 {\bar 3}}{\langle 2 {\bar 1}\rangle}\right) W_3\right]
 \label{eq:TLLunfixed}
\end{eqnarray}
for some constants ${\hat\lambda}_{0,1}$.    (A possible term proportional to $z^2$ times the symmetric invariant in the first line of this equation vanishes, as can be seen in a frame with $Z_3$ at the origin and $Z_2$ at infinity).     To fix the constants, we impose Ward identities.   Taking $x_1=0$ and $x_2\rightarrow\infty$, the $\theta_3 {\bar\theta}_3$ component of this equation is
\begin{equation}
\langle T_{\mu\nu}(x_3) C^I(0) C^J(x_2\rightarrow\infty)\rangle = {2 \delta^{IJ}\over x_2^4 x_3^4} \left[{\hat\lambda}_0 \eta_{\mu\nu} + 2 {\hat\lambda_1} {{x_3}_\mu {x_3}_\nu\over x_3^2}\right].
\end{equation}
Comparing this to the energy-momentum Ward identity~\cite{Callan:1970ze}
\begin{equation}
\partial_3^\nu \langle T  T_{\mu\nu}(x_3) C^I(0) C^J(x_2\rightarrow\infty)\rangle  = -i \partial^3_\mu\left[ \delta^4(x_3-x_1) +\delta^4(x_3-x_2)\right]\langle T C^I(x_1) C^J(x_2)\rangle
\end{equation}
requires that ${\hat \lambda}_1 = - 2 {\hat \lambda_0}$ and ${\hat\lambda}_0 = \tau/(2^6 \pi^6)$.   This yields
\begin{eqnarray}
\nonumber
\langle L^I(Z_1) L^J (Z_2) \left[{\bar W}_3 \mathcal{T} (Z_3)W_3 \right]  \rangle &=&  {\tau\delta^{IJ} \over 64\pi^6\langle 3 {\bar 1} \rangle \langle1 {\bar 3}\rangle   \langle 3 {\bar 2} \rangle \langle 2 {\bar 3}\rangle}\ {\bar W}_3\left [(1-2z) \frac{3 {\bar 2} 1 {\bar 3}}{\langle 1 {\bar 2}\rangle}\right. \nonumber \\
& &{} \left.  + (1+2 z)\frac{3 {\bar 1} 2 {\bar 3}}{\langle 2 {\bar 1}\rangle}\right] W_3 \, .
\end{eqnarray}
Because this correlator also contains the term $\langle j_{\mu_1}^R j_{\mu_2}^I j_{\mu_3}^J\rangle$,  we recover the relation between the normalization of the global current two-point function and the mixed $\mbox{Tr} RG^2$ anomaly of the SCFT.

As our last application, we consider the correlator $\langle {\mathcal T}  {\mathcal T} L \rangle$. The form of tensors that appear in this correlator is restricted by the ``gauge invariance" of Eqs.~(\ref{eq:gaugeW}) and (\ref{eq:gaugeWbar}) to be of the form 
\begin{eqnarray}
\nonumber
\left[{\bar W}_1  (1 \ldots \bar{1} ) W_1\right] \cdot  \left[{\bar W}_2  (2 \ldots \bar{2} ) W_2\right],\\
\nonumber
\left[{\bar W}_1  (1 \ldots \bar{2} ) W_2\right] \cdot  \left[{\bar W}_2  (2 \ldots  \bar{1} )W_1\right],\\
\left[{\bar W}_1  (1\ldots 2 )  \bar{W}_2 \right] \cdot  \left[ W_1(  \bar{1} \ldots \bar{2})  W_2\right].
\end{eqnarray}
times powers of $z$ and $z^2$.   The omitted expressions, denoted by $(\ldots)$, are strings of products of supercoordinates $Z_{1,2,3}$ which are constructed by imposing  $SU(2,2|1)$ invariance together with the covariant constraints Eq.~(\ref{constraints16}) and Eq.~(\ref{constraints24}).  The $[X \bar{X}]_{\bf{24}}=0$ and $[X \bar{X}]_{\bf{1}}=0$  constraints imply that such strings  cannot contain a coordinate times its conjugate in adjacent positions.
The $[X X]_{\bf{16}}=0$ constraint written as in Eq.~(\ref{16}) is useful for rearranging/shortening strings with repeated insertions of a given coordinate.  As a specific example, it follows readily from Eq.~(\ref{16}) that 
\begin{equation}
 (1{\bar 2}1)_{AB} = \frac{1}{2}\langle 1{\bar 2}\rangle {X_1}_{AB}.
\end{equation}
Using the constraints in this way to simplify possible tensors, the  $\langle {\mathcal T}  {\mathcal T} L \rangle$ correlator reduces to six structures with definite reality and permutation symmetry properties:
\begin{eqnarray}
t_1 &=&  \frac{\left[{\bar W}_1 (1 \bar{2}) W_2\right] \,  \left[{\bar W}_2 (2 \bar{1}) W_1\right]}{\left\langle 1\bar{2}\right\rangle\left\langle 2\bar{1}\right\rangle}, \\
t_2  &=&  \frac{\left[{\bar W}_1 (1 \bar{3} 2) \bar{W}_2 \right] \,  \left[ W_1 (\bar{1} 3 \bar{2}) W_2\right]}{\left(\left\langle 1\bar{2}\right\rangle\left\langle 2\bar{1}\right\rangle\left\langle 1\bar{3}\right\rangle\left\langle 3\bar{1}\right\rangle\left\langle 2\bar{3}\right\rangle\left\langle 3\bar{2}\right\rangle\right)^{\frac{1}{2}}}, \\
t_3 &=&  \frac{\left[ {\bar W}_1 (1 \bar{2} 3 \bar{1})  W_1\right] \,  \left[{\bar W}_2 (2 \bar{1} 3 \bar{2}) W_2\right]}{\left\langle 1\bar{2}\right\rangle\left\langle 2\bar{1}\right\rangle\left\langle 3\bar{1}\right\rangle\left\langle 3\bar{2}\right\rangle}
+  \frac{\left[{\bar W}_1 (1 \bar{3} 2 \bar{1}) W_1\right] \,  \left[{\bar W}_2 (2 \bar{3} 1 \bar{2}) W_2\right]}{\left\langle 1\bar{2}\right\rangle\left\langle 2\bar{1}\right\rangle\left\langle 1\bar{3}\right\rangle\left\langle 2\bar{3}\right\rangle} ,  \\
t_4 &=& \frac{\left[ {\bar W}_1 (1 \bar{2} 3 \bar{1})  W_1\right] \,  \left[{\bar W}_2 (2 \bar{1} 3 \bar{2}) W_2\right]}{\left\langle 1\bar{2}\right\rangle\left\langle 2\bar{1}\right\rangle\left\langle 3\bar{1}\right\rangle\left\langle 3\bar{2}\right\rangle}-   \frac{\left[{\bar W}_1 (1 \bar{3} 2 \bar{1}) W_1\right] \,  \left[{\bar W}_2 (2 \bar{3} 1 \bar{2}) W_2\right]}{\left\langle 1\bar{2}\right\rangle\left\langle 2\bar{1}\right\rangle\left\langle 1\bar{3}\right\rangle\left\langle 2\bar{3}\right\rangle}  , \\
t_5 &=& \frac{\left[ {\bar W}_1 (1 \bar{2} 3 \bar{1})  W_1\right] \,  \left[{\bar W}_2 (2 \bar{3} 1 \bar{2}) W_2\right]}{\left\langle 1\bar{2}\right\rangle^2\left\langle 3\bar{1}\right\rangle\left\langle 2\bar{3}\right\rangle}
+  \frac{\left[{\bar W}_1 (1 \bar{3} 2 \bar{1}) W_1\right] \,  \left[{\bar W}_2 (2 \bar{1} 3 \bar{2}) W_2\right]}{\left\langle 2\bar{1}\right\rangle^2\left\langle 1\bar{3}\right\rangle\left\langle 3\bar{2}\right\rangle}  , \\ 
t_6 &=& \frac{\left[ {\bar W}_1 (1 \bar{2} 3 \bar{1})  W_1\right] \,  \left[{\bar W}_2 (2 \bar{3} 1 \bar{2}) W_2\right]}{\left\langle 1\bar{2}\right\rangle^2\left\langle 3\bar{1}\right\rangle\left\langle 2\bar{3}\right\rangle}
-  \frac{\left[{\bar W}_1 (1 \bar{3} 2 \bar{1}) W_1\right] \,  \left[{\bar W}_2 (2 \bar{1} 3 \bar{2}) W_2\right]}{\left\langle 2\bar{1}\right\rangle^2\left\langle 1\bar{3}\right\rangle\left\langle 3\bar{2}\right\rangle}.    
\end{eqnarray}
Not all these objects are independent.  By going to the special frame where $Z_{2,3}$ are fixed, and covariantizing the result, it is possible to establish certain relations between the structures given above. These relations are $z t_4=0$, $z t_6= -z^2 t_3$ , $t_3+t_5=- z^2 t_2$, and  $z^2(t_3-t_5) = z^2(t_1+2 t_2)$.    Thus, the most general form of the three-point function   $\langle {\mathcal T}  {\mathcal T} L \rangle$ can be taken to be of the form 
\begin{eqnarray} 
\nonumber
\label{eq:ttl}
  \langle \left[{\bar W}_1{} \mathcal{T} (Z_1)W_1 \right] \,\left[ {\bar W}_2 \mathcal{T}(Z_2) {W_{2}} \right]\,  L(Z_3)\rangle =  {(\lambda_1+ \tilde{\lambda}_1z^2)  t_1 + \lambda_2 t_2 + \lambda_4 t_4 + (\lambda_5 + \tilde{\lambda}_5 z^2) t_5 \over  \left[\left\langle 1\bar{2}\right\rangle\left\langle 2\bar{1}\right\rangle\left\langle 1\bar{3}\right\rangle\left\langle 3\bar{1}\right\rangle\left\langle 2\bar{3}\right\rangle\left\langle 3\bar{2}\right\rangle\right]^{1/2}}.\\
\end{eqnarray}

The coefficients in this expression are related by conservation of the supercurrent, $D^a T_{a{\dot a}}={\bar D}^{\dot a} T_{a{\dot a}}=0$.   In the special frame with $X_2\rightarrow 0$, $X_3\rightarrow\infty$, Eq.~(\ref{eq:ttl}) contains the component correlation function
\begin{eqnarray}
\langle j^R_{\mu}(x) j^R_{\nu}(0) C(\infty)\rangle = {1\over x^4 x^4_3} \left[- \eta_{\mu \nu} \left(\lambda_1 -{1\over 2}\lambda_2\right) + {x_{\mu} x_{\nu}\over x^2} (2 \lambda_1 - \lambda_5)\right].
\end{eqnarray}
Conservation of the $R$-current, $\partial^\mu j^R_\mu=0$  then fixes $\lambda_5= 2 (\lambda_2-\lambda_1)$.   In this frame, the correlator involving the energy-momentum tensor is
 \begin{eqnarray}
\langle T_{\mu \nu}(x) j^R_{\rho}(0) C(\infty)\rangle = {1\over x^6 x^4_3} \left[ \lambda_1 \epsilon_{\mu \nu  \rho \alpha} \, x^\alpha - \frac{i}{2}\lambda_4 \left(x_{\mu} \eta_{\nu \rho}- x_{\rho} \eta_{\nu \mu }\right)\right],
\end{eqnarray}
so that symmetry $T_{\mu \nu} = T_{\nu \mu}$ requires $\lambda_1=\lambda_4=0$.    Finally, the $\theta^2{\bar\theta}^2$ component of the superfield $\mathcal{T}_{a{\dot a}}(x,\theta,{\bar\theta})$ vanishes, which implies that ${\tilde\lambda}_1=-{\tilde\lambda}_5=- \lambda_2$.  Combining together the constraints from different components, we obtain 
\begin{eqnarray} 
&&  \langle \left[{\bar W}_1{} \mathcal{T} (Z_1)W_1 \right] \,\left[ {\bar W}_2 \mathcal{T}(Z_2) {W_{2}} \right]\,  L(Z_3)\rangle = 
 \frac{c_{TTL}}{\left(\left\langle 1\bar{2}\right\rangle\left\langle 2\bar{1}\right\rangle\left\langle 1\bar{3}\right\rangle\left\langle 3\bar{1}\right\rangle\left\langle 2\bar{3}\right\rangle\left\langle 3\bar{2}\right\rangle\right)^{\frac{1}{2}}} \nonumber \\
&&\; \times \left[  z^2 \frac{\left[{\bar W}_1 (1 \bar{2}) W_2\right] \,  \left[{\bar W}_2 (2 \bar{1}) W_1\right]}{\left\langle 1\bar{2}\right\rangle\left\langle 2\bar{1}\right\rangle} 
 -  \frac{\left[{\bar W}_1 (1 \bar{3} 2) \bar{W}_2 \right] \,  \left[ W_1 (\bar{1} 3 \bar{2}) W_2\right]}{\left(\left\langle 1\bar{2}\right\rangle\left\langle 2\bar{1}\right\rangle\left\langle 1\bar{3}\right\rangle\left\langle 3\bar{1}\right\rangle\left\langle 2\bar{3}\right\rangle\left\langle 3\bar{2}\right\rangle\right)^{\frac{1}{2}}} \right.  \nonumber \\
&& \left. - (2+ z^2)\left(  \frac{\left[ {\bar W}_1 (1 \bar{2} 3 \bar{1})  W_1\right] \,  \left[{\bar W}_2 (2 \bar{3} 1 \bar{2}) W_2\right]}{\left\langle 1\bar{2}\right\rangle^2\left\langle 3\bar{1}\right\rangle\left\langle 2\bar{3}\right\rangle}
+  \frac{\left[{\bar W}_1 (1 \bar{3} 2 \bar{1}) W_1\right] \,  \left[{\bar W}_2 (2 \bar{1} 3 \bar{2}) W_2\right]}{\left\langle 2\bar{1}\right\rangle^2\left\langle 1\bar{3}\right\rangle\left\langle 3\bar{2}\right\rangle}    \right)  \right].
\end{eqnarray}
Therefore, superconformal invariance determines the $\langle {\mathcal T}  {\mathcal T} L \rangle$ correlator up to the overall normalization, which is in agreement with Ref.~\cite{Osborn:1998qu}.

\section{Conclusions}
\label{Conclusion}

In this paper, we have shown how superconformal multiplets in representations $(j,{\bar j},q,{\bar q})$ fit into the superembedding framework introduced in~\cite{GSS}.  Physically, the most important examples  correspond to the real scalar multiplet that contains the global conserved current $j_\mu$ and the supercurrent multiplet $\mathcal{T}_{a\dot{b}}$ containing $j^\mu_R$, the supercurrent $S^\mu_a$   and the energy momentum tensor ${T^\mu}_\nu$.   Constructing the relevant correlators is reduced to the task of enumerating $SU(2,2|1)$ invariants that appear in the products of several copies of the  linear representations  $X_{AB}$, ${\bar X}^{AB}$, $W_A$, ${\bar W}^A$.    This index-free approach yields relatively compact expressions for the Green's functions.   Although the examples presented are not new, the manifestly covariant forms we presented are, and we hope that the simplifications that come with working in the superembedding formalism will eventually lead to new results.

At present we have no way of representing contact terms in the embedding formalism.    This would be necessary, for instance, to deal with the anomaly structure of conserved current three-point functions in a covariant way (rather than imposing that the formalism satisfies the correct anomaly relations component-wise, as we did in this paper).    To this end, a direct embedding space formulation of conservation laws such as $D^2 L={\bar D}^2 L=0$, etc., in terms of embedding space differential operators would also be required.  Finally, besides the extension of the formalism discussed here to the case of extended superconformal invariance, another useful directions might be to see if the recent techniques developed in~\cite{SimmonsDuffin:2012uy} for efficiently computing conformal blocks have a natural extension to the supersymmetric case.

\centerline{\bf Acknowledgements}
\noindent We thank Francesco Iachello and Minho Son for discussions. This work is supported in part by DOE grant DE-FG-02-92ER40704. 
WS thanks the KITPC and the Aspen Center for Physics for their hospitality where parts of this work have been done, and for partial support, respectively, by the Project of Knowledge Innovation Program (PKIP) of Chinese Academy of Sciences, Grant No. KJCX2.YW.W10 and  by the National Science Foundation under Grant No. PHYS-1066293. 


\end{document}